# PROBABILISTIC VERIFICATION AND EVALUATION OF BACKOFF PROCEDURE OF THE WSN ECo-MAC PROTOCOL

Hafedh ZAYANI[1] and Kamel BARKAOUI[2] and Rahma BEN AYED[1]

[1] National Engineering School of Tunis (ENIT), SysCom, BP. 37 le Belvedere 1002 Tunis, Tunisia
{hafedh.zayani, rahma.benayed}@enit.rnu.tn

[2] CNAM, CEDRIC, Paris, 292, Rue Saint-Martin 75141 Paris Cedex 03
kamel.barkaoui@cnam.fr

## ABSTRACT

*Communication protocols and techniques are often evaluated using simulation techniques. However, the use of formal modeling and analysis techniques for verification and evaluation in particular for Wireless Sensor Networks (WSN) becomes a necessity. In this paper we present a formal analysis of the backoff procedure integrated in the medium access control protocol named ECo-MAC designed for WSN. We describe this backoff procedure in terms of discrete time Markov chains (DTMCs) and evaluated using the well known probabilistic model checker PRISM. After checking the different invariants of the proposed model, we study the effect of contention window length (in number of time contention unit) on the acceptable number of simultaneous senders in a neighborhood of a given receiver. The obtained quantitative results confirm those provided by the simulation using OPNET tool and justify the validity of the adopted value for the time contention unit TCU.*

## 1. INTRODUCTION

Wireless sensor networks (WSN) is composed of many sensor nodes, scattered throughout a target area, that monitor some events witch depend of the application characteristics. Collected data by these sensor nodes has to be forwarded to a destination node called Base Station (BS) [1]. WSNs have been an active research topic during the past few years because they have been proposed for a large range of application and they have some constraints such as limited battery lifetime, reduced memory capacity, etc. Researchers have mainly conducted extensive studies on energy saving and other performance criteria using the simulation techniques and discrete event simulators.

Constraints induced by the wireless medium motivate the implementation of new communication protocols characterized by a probabilistic behavior. This characteristic cannot provide complete coverage of a model proposed using simulation environment. To overcome these limitations, model checker techniques are proposed [15]. These techniques can be used to provide an exhaustive analysis and an automatic verification of partial or abstract model. Therefore, model checker requires the use of abstracted or simplified model that must be true to the real system behavior. Several frameworks can facilitate modeling and checking process such as UPPAAL (Uppsala University and Aalborg University project) [17], SPIN/PROMELA (Simple PROMELA INterpreter/PROcess MEta LAnguage) [19], PRISM (PRobabilistIc Symbolic Model checker) [18], etc.

In wireless communication, several medium access control protocols use a backoff procedure in retransmission phase to reduce the likelihood of reappearance of a collision. Different procedures are described and proposed such as the Binary Exponential Backoff (BEB) used in the international standard IEEE 802.11 [3], the normal distribution introduced by Gobriel et al. in [2], the Impatient Backoff Algorithm (IBA) studied in [4], the Multiplicative Increase Linear Decrease (MILD) [5], the Double Increase Double Decrease (DIDD) [6], etc. In WSN we had proposed a new medium access control protocol ECo-MAC [7] with a new backoff procedure. We had studied the effectiveness of the proposed procedure using OPNET [16] simulation.

Different works have studied a BEB backoff procedure by the simulation technique [8], by the analytic approach [9] and using the UPPAAL and PRISM model checkers [10]. In this paper, we consider automatic verification and quantitative evaluation of the backoff procedure of ECo-MAC protocol implemented in OPNET simulator. For probabilistic modeling and analysis, we have used the probabilistic model checker PRISM framework. This tool has proven to be successful in a wide range of case studies [10], [11], [12], [13], [14].

The paper is organized as follows. We begin by presenting the proposed backoff mechanism with the ECo-MAC protocol in section 2. Then, we describe the proposed probabilistic model of this procedure in section 3 using PRISM framework. Finally, we analyze the proposed model with PRISM tool in section 4, 5 and 6. We terminate the paper by a conclusion in the section 7.

## 2. BACKOFF PROCEDURE USED IN ECO-MAC PROTOCOL

As described in [7], ECo-MAC is a media access control protocol for wireless sensor networks. This protocol is hybrid. It takes advantage of three access techniques CSMA, TDMA (Time Division Medium Access), and multi-channel protocols. Differently of classical TDMA protocols that manipulate two level of time division (frame and time slot), ECo-MAC uses only one level of division into different time slots (TS). All time slots have a constant length that is depend on radio TX/RX characteristics and different protocol parameters. Considering that the majority of WSN applications are characterized by a low traffic, authors of ECo-MAC decide that at a beginning of each TS, all nodes having data to transmit send immediately an RTS without listening to the channel. This makes all nodes, except the destination addressed in the RTS to switch to the SLEEP mode for the remaining time in the currents TS. Two situations may occur as an RTS overcome. (1-success) The source receives the expected CTS from the destination node and then the communication takes place within a sub-band frequency randomly selected by the source node and transmitted in the RTS packet. (2-fail) The source node receives a JAM packet or the receiving response timer expires without receiving a valid CTS frame, meaning that a collision occurs. To avoid collision in the retransmission of current data packet, node executes a backoff procedure to reduce a possibility of collision in the retransmission phase in current TS.

Sender nodes that fail the free access in the first sending phase, must execute a backoff procedure in a second sending phase in the current TS. In contention phase, each deferring node chooses backoff duration composed of an integer random number ($rbc$) of time contention units ($T_{CU}$). Value of $rbc$ is drawn from a uniform distribution $uRand$ over the interval $[b_i, b_f]$. These bounds $b_i$ and $b_f$ represent respectively the low and the upper limits integer value of the allowed number for the backoff counter. These bounds depend on the consecutive unsuccessful transmissions number ($e$) of a current packet. An unsuccessful transmission is represented by fail sending in the current TS using two attempts. The value of $b_i$ has a lower bound of zero and the upper value of $b_f$ is a number ($b_{max}$) related to the application requirements and characteristics. Also, the value of e has an upper bound of $e_{max}$ that is related to the maximum end-to-end latency allowed and the TS duration. If unsuccessful transmissions number of current packet reaches the upper bound, the packet will be rejected. Where there is not activity in the channel for the $T_{CU}$ duration, the backoff counter $rbc$ is reduced by one and when the $rbc$ reaches the zero value, the correspondent node must starts its transmission. But, if the medium becomes busy the reduction of $rbc$ is ignored and the node delays its transmission to the next TS and it increases the unsuccessful transmissions number ($e$) by one. In the next TS, the node that loses contention, not uses the last value of $rbc$ but it must chooses a new value in a new interval that depends on $e$ value. More formally, the backoff procedure is based on the function described in Equation 1.

$$B : [0..e_{max}] \rightarrow [0..b_{max} * T_{CU}]$$
$$e \mapsto B(e) = rbc * T_{CU}$$
$$Where :$$
$$\bullet \ rbc = uRand[b_i..b_s] \ \ \forall e \in [e_i..e_s]$$
$$\bullet \ [b_i..b_s] \subset [0..b_{max}] \ and \ [e_i..e_s] \subset [0..e_{max}]$$

(1)

In figure 1, the nodes $n1$ and $n3$ want to send data towards the same neighbor $n_2$. If node $n_1$ win contention (i.e. $n_1$ and $n_3$ have chosen different backoff time), the node $n3$ must realizes that $n_1$ has won current contention before that the $n_3$ decrease its backoff counter $rbc$ to reduce the likelihood collision caused by the hidden nodes problem. This observation imposes a constraint on the contention unit duration $T_{CU}$ value. This value must equal at least to the RTS transmission time and the time that allows $n_2$ starts to send its response CTS. Equation 2 shows a formulation of the $T_{CU}$ value, where $T_{MxSRT}$, $T_{FrmCtrl}$ and $T_{RSSI}$ are defined respectively by maximal time switching between TX and RX, sending time of a control frame and Received Signal Strength Indicator (RSSI) time.

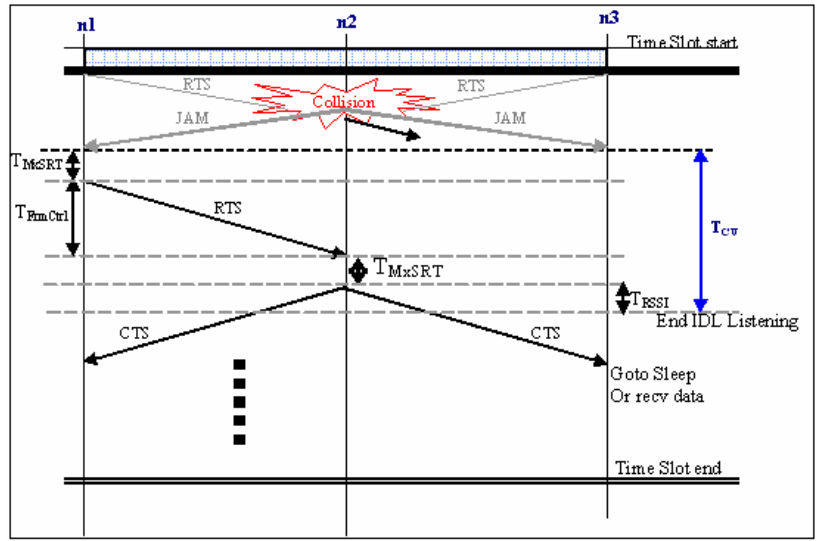

Figure 1. Modelling Backoff procedure with PRISM model checker

$$T_{CU} = 2 * T_{MxSRT} + T_{FrmCtrl} + T_{RSSI}$$

(2)

In the remaining of this paper we study the backoff procedure as described in Equation 3.

$$B : [0..12] \rightarrow [0..7 * T_{CU}]$$

$$e \mapsto B(e) = \begin{cases} uRand[1..7] * T_{CU} \ if \ e \in [0..1] \\ uRand[0..7] * T_{CU} \ if \ e \in [2..3] \\ uRand[0..6] * T_{CU} \ if \ e \in [4..6] \\ uRand[0..5] * T_{CU} \ if \ e \in [7..8] \\ uRand[0..4] * T_{CU} \ if \ e \in [9..10] \\ uRand[0..3] * T_{CU} \ if \ e \in [11..12] \end{cases} \quad (3)$$

## 3. MODELING BACKOFF PROCEDURE WITH PRISM MODEL CHECKER

In this section, we start by an overview of the PRISM model checker tool that we have been used in modelling and analyzing process. In this overview, we summarize the supported models, the modeling language, and the specific temporal logics language used to express some properties of DTMC model. Next we describe the network configuration used for probabilistic modeling phase. Then, we describe each component of entire proposed probabilistic model.

### 3.1. Network Configuration

For all studied scenarios, we consider a fixed network consisting of a one receiver and a number of senders (sender1, sender2, etc.). The choice of this topology that has a hidden sender nodes is motivated by the need to verify the validity of the chosen $T_{CU}$ duration, and to know the effect of varying the number of sender in the same neighborhood on the successfully sending packet probability. Figure 2 show an example based on two senders. Each sender is within communication range of the center receiver, but the senders cannot communicate with each other, as they do not have a physical connection to each other. These senders are known as hidden nodes. Our probabilistic model concerns not all the ECo-MAC protocol but only its backoff procedure. Then, each sending node intends to transmit, using backoff procedure of ECo-MAC, a data frame to the central receiver. If both senders start their transmission at the same time, it presupposes sending nodes have chosen a same backoff counter rbc value, and then a collision must be occurred and detected by the receiver. Consequently, the receiver doesn't answer by a CTS (Clear To Send) frame, if it has a collision between both RTS (Request To Send) request sent by senders. Afterwards, the senders detect this collision after waiting response during a given timeout, and without receive valid CTS.

If senders have chosen different backoff counter *rbc* values, then the one that has chosen the lowest value, wins the current contention and it starts its transmission before the others. Consequently, receiver must respond the received valid RTS by CTS toward the winner sender. When, the receiver starts sending CTS, hidden nodes detect contention lost during the current contention unit $T_{CU}$ and before they decrease their backoff counters *rbc*. In that case, these nodes reset their backoff counters *rbc*, increase their unsuccessful transmission counters, and delay their transmissions to the next time slot TS.

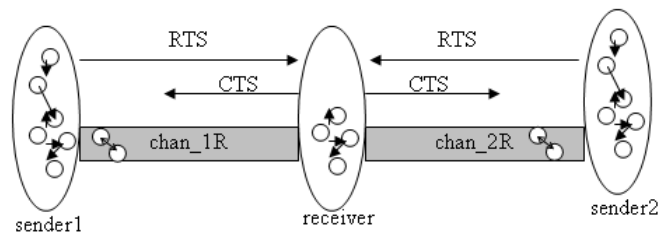

Figure 2. Two hidden sender nodes scenario

We would study formally the backoff procedure implemented with the ECo-MAC [7] protocol. For that, we have developed a small, but a detailed, probabilistic DTMC model of this procedure using the model checking PRISM tool. The model comprises several modules: one for the receiver node in the network and one representing a channel connecting a couple of nodes the order in which nodes executes the backoff procedure to send their data frame. Then, we will be describing the following basic modules: receiver, chan_1R, and sender1; to add, for example, other sender (*sender₂*), we can use the module renaming technique for the defined sender1 module. We have used a graphical representation using the finite state machine (FSM) formalism that is not PRISM's representation, to describe better the behavior of each module.

### 3.3. Receiver module

The receiver waits possibly RTS request from one of the senders during $< (1 + b_{max}) * T_{CU} >$. When the medium becomes busy, the receiver receives the current frame and decides if there is a collision or not. The receiver aborts the collided transmission with no response, and it sends CTS response after receiving a valid RTS request from one receiver.

- − W_START: Waits the beginning of contention phase

- − W_RTS: Waits RTS frame

- − COLLISION: detects collision between two different RTSs frames

- − RCV_F_S1: receives RTS from sender1

- − RCV_F_S2: receives RTS from sender2

- − W_S_RT: Waits until the radio switches between RX to TX mode.

- − W_S_CTS: Waits until the CTS frame have been sent (sending time)

- − W_END: Waits until both senders detect collision occurred on the current transmission.

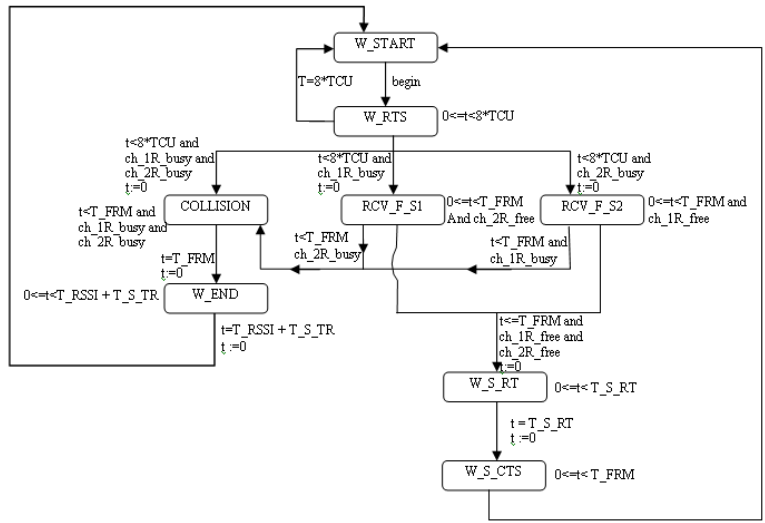

Figure 3. receiver's behavior

### 3.4. Channel module

This module describes the wireless channel that simulates the physical connection between nodes. Module chan_1R specifies the channel that is connect sender1 to the receiver. Behavior of this module can be represented by the FSM of figure 4. The channel changes its state according to the decision and states of connected nodes. The FSM has the following states:

– IDLE: the channel is free (all connected nodes are in inactive mode).

– COLLISION: collision occurs between at least tow simultaneous transmissions.

– BUSY_RECEIVER: Only the receiver node is in sending mode.

– BUSY_SENDER1: Only the sender1 node is in sending mode.

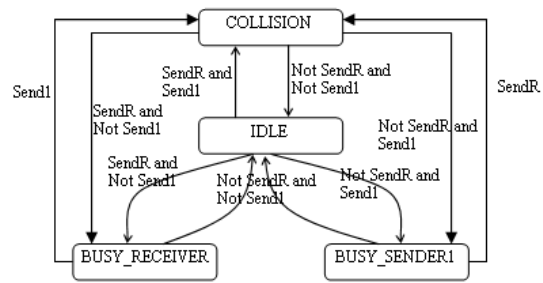

Figure 4. chan1R's behavior

### 3.4. Sender module

This module describes the behavior of the sender node in contention phase using backoff procedure implemented with ECo-MAC protocol. The sender node starts by choosing a random backoff counter *rbc* value. Next, it listens to the medium during *rbc* * $T_{CU}$. If this duration terminates with a free channel, the corresponding sender sends its RTS request. If not, the node switches to sleep state waiting end of current transmission. Figure 5 presents the FSM describing behaviour of sender1. As shown in this figure, sender1 starts by choosing a random backoff counter *rbc* within the first contention window associated to the unsuccessful transmissions number $e = 0$. This state is notated in the FSM by the couple of values ($e = 0$; *rbc*). For each value of the *rbc* that is different from zero, the sender listen the channel during one $T_{CU}$.

If the medium remains free over the current TCU, the sender decreases the backoff counter *rbc* value by one. When the *rbc* value reaches zero, the node begins its transmission by sending RTS request frame, after switching from radio RX to TX mode. After sending RTS frame, it switches from radio TX to RX mode, and listen the channel during RSSI time. When the node receives a valid CTS frame, it assumes that the current packet is sent correctly and it reset e and *rbc* values for a new transmission. If the RSSI time expires without receiving a valid CTS frame, the sender assumes that is a collision occurs. For that, the node increases the e value by one and it resets the *rbc* value by a new chosen value within a new contention windows considering the new value of *e*.

If the medium become busy before the current $T_{CU}$ expires, the sending node aborts its currents listening. In this case, the node increases the e value by one and it resets the *rbc* value by a new chosen value within a new contention windows relating to the new value of $e$. When e reaches the $e_{max}$ value, the node rejects the current packet and it resets the e value by zero for the next packet.

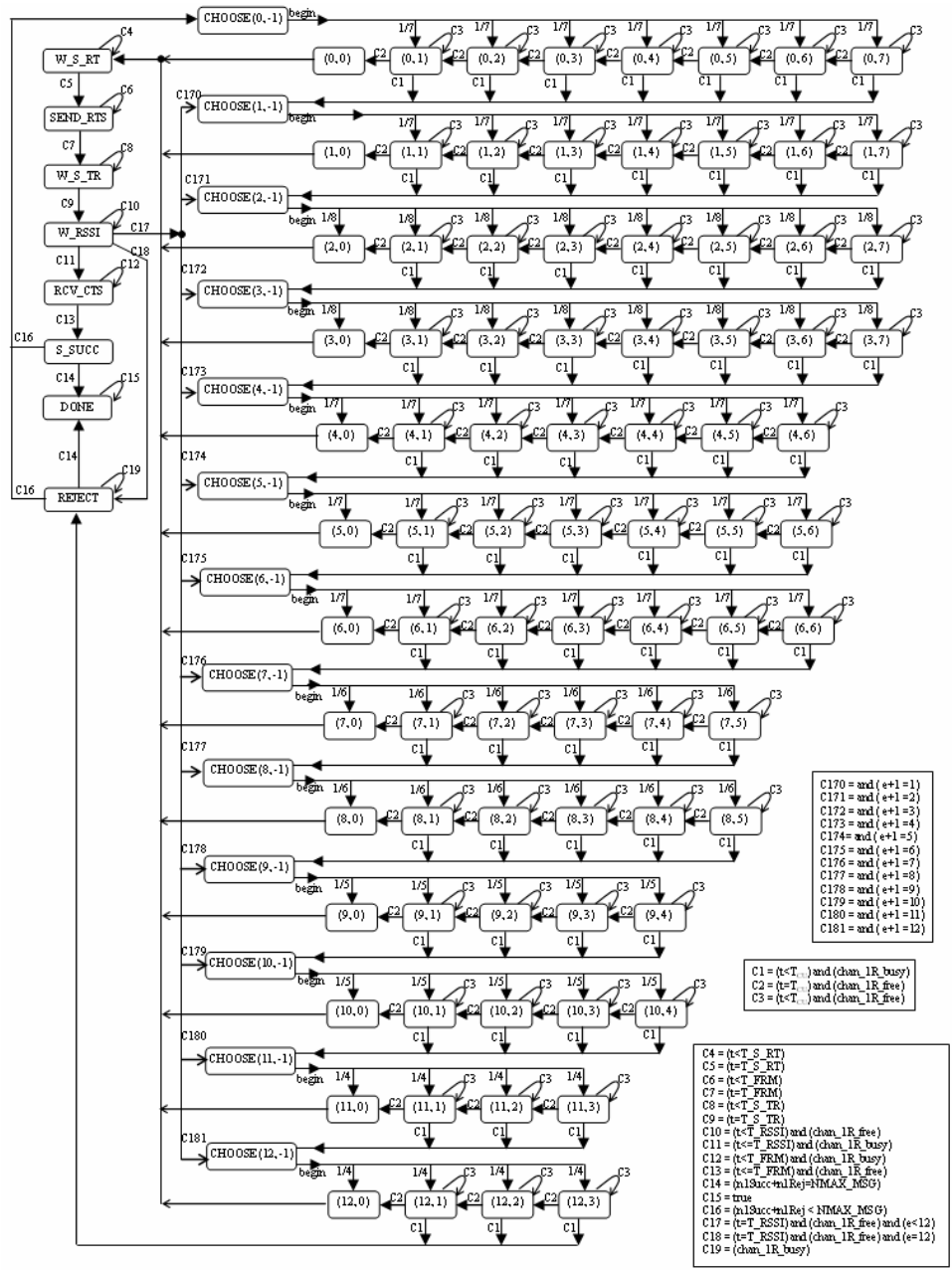

Figure 5. sender1's behaviour

The FSM in figure 5 consists of the following states:

- CHOOSE ($e$, -1): Chooses a random backoff counter $rbc$ value from a contention window relating with e value.

‒ (*e*, *rbc*): gives the backoff counter value for each time depending on the transmission history of the current packet.

‒ W_S_RT: Waits until the radio switches from radio RX to TX mode.

‒ SEND_RTS: Waits until the RTS frame have been sent (RTS sending time)

‒ W_S_TR: Waits until the radio switches from radio TX to RX mode.

‒ W_RSSI: RSSI time.

‒ RCV_CTS: Waits until the CTS frame have been received (CTS receiving time).

‒ S_SUCC: the current packet is sent correctly (successful of the current transmission)

‒ DONE: final state, when the sender sent all packets.

‒ REJECT: rejects the current packet.

## 4. CONFORMITY AND VALIDITY OF THE PROPOSED PROBABILISTIC MODEL

The ECo-MAC protocol is implemented using the professional OPNET [16] simulator. The modeled scenario presented on figure 2 using probabilistic model checking PRISM tool, is proposed for only the contention phase transmission based on the backoff procedure of the ECo-MAC protocol. To study the conformity of the proposed probabilistic model, we have opted for making a comparison between OPNET and PRISM results. This comparison is base on the idle listening time introduced by the backoff procedure evaluated after that each sender sends NMAX MSG packets in both tools. For that, we have reproduced the scenario, presented in figure 2, in OPNET simulator, where each sender (resp. receiver) is replaced by a sensor (resp. base station) node. For each value of NMAX MSG, the simulation was repeated 10 times and the results present confidence interval of 95%. In order to evaluate this time in PRISM model checker, we have used the reward structure. The reward structure must assign the TCU duration to each transition between two couple (*e*, *rbc*). Table I gives the values of different parameters used in both tools OPNET and PRISM.

Table 1. Values of the parameters used in OPNET and PRISM

| Parameters [units] | Values |
|---|---|
| Band-width [bytes/s] | 10000 |
| W_S_TR [µs] | 850 |
| W_S_RT [µs] | 850 |
| RTS length [bytes] | 14 |
| CTS length [bytes] | 14 |
| T_RSSI [µs] | 12 |
| NMAX_MSG | 1..5 |

The proposed DTMC model in PRISM model checker, time is discrete and is bounded. In a first version of the probabilistic model, we have used the same microsecond ($\mu s$) unit for all parameters. With this choice, the $T_{FrmCtrl}$ and $T_{CU}$ parameters values will be equal to 12000 and 13712 microseconds ($\mu s$) respectively. These great parameter values involve increasing the state space resulted in construction of the model that is terminated by state space explosion problem.

This problem become after more than four hours of model building. To avoid this problem, we have reduced all values and we keep the percentage reduction of each parameter that was been used in the reward structures. Equation 4 illustrates the reward structure named *TidleS1*. This reward is used to evaluate the idle listening time introduced by the backoff procedure of the *sender1*. The reward structure takes into a count of a percentage reduction. This value is the ratio of the modified $T_{CU}$ value in PRISM model (8 units) to the realistic $T_{CU}$ value (13712 $\mu s$). Equation 5 shows the PCTL formulation of the reward property. This property is used to find results presented in figure 6.

$$rewards \quad " \, TidleS1 \, "$$
$$ss1 = 0 \; : \; 0.001714;$$
$$endrewards \tag{4}$$

$$R_{"TidleS1"=?}[F(ss1 = 9)] \qquad NMAX\_MSG \in [1..5] \tag{5}$$

In Figure 6, we have plotted the idle listening time introduced by the backoff procedure of the sender1 for different values of the NMAX_MSG parameter using PRISM model checker and OPNET simulator. We notice that the results obtained with PRISM model checker is always slightly greater compared to the OPNET simulator results. We will be able to justify this light difference between the founded results, as follow. The proposed probabilistic model is partial, but the OPNET model implements a complete ECo-MAC protocol. As it presented in section 3, the sending node, executing ECo-MAC protocol, has a first phase with free contention at each beginning of TS. And if the transmission is failed, the sender executes a backoff procedure in the second phase of the current TS. Then, if sender1 terminates sending their NMAX_MSG packets before the other, in PRISM the sender2 continues sending their remaining packets using backoff procedure but in OPNET the node sends these packets during the first phases and then without contention. Therefore, the time that will be obtained with PRISM must be greater than that obtained with OPNET.

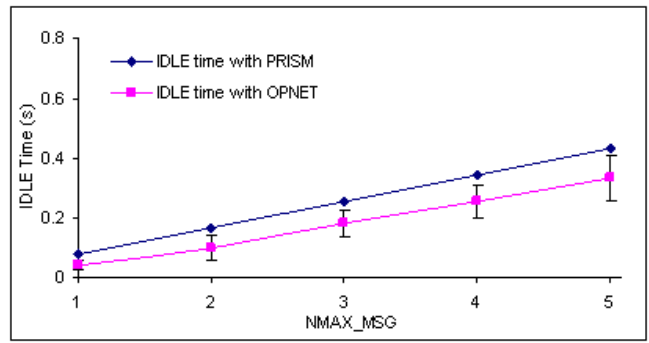

Figure 6. Idle listening time introduced by the backoff procedure of the *sender1*

In order to verify the validity of the proposed probabilistic backoff model, we have verified different properties. The first property consists in checking that if sender1 rejects its packet then the unsuccessful transmissions $e$ number has been reached the bounded value $e_{max}$ (12). Equation 6 shows the formulation of this property using PCTL logic language, where ($ss1$=8) stands for the REJECT state in the *sender1* module in PRISM and es1 is a local variable that stores the value of $e$ for the sender1. Checking of this property in PRISM model checker returns true value.

$$P_{>=1}[G(ss1 = 8) => (es1 = 12)] \tag{6}$$

The second property tends to verify that if sender1 have sent successfully the current packet, it must has the value of unsuccessful transmissions e number in the range [0..12]. The PCTL formulation of this property is given by the Equation 7, where ($ss1$=7) refers to the S SUCC state in *sender1* module. When this property is checked, PRISM model checker returns true value.

$$P_{>=1}[G\,(ss1 = 7) => (es1 >= 0)\&(es1 <= 12)] \qquad (7)$$

The third property verifies if there is a possibility that the sender1 rejects its current packet, and it has a value of unsuccessful transmissions e number less then 12. The PCTL formulation of this invariant is given in Equation 8. The property is not satisfied by the model checker PRISM, in fact each sending node will not be able to reject current packet before that the corresponding e number reaches the $e_{max}$ value.

$$P_{>=1}[G\,(ss1 = 8) => (es1 >= 0)\&(es1 < 12)] \qquad (8)$$

We would verify that if the two hidden senders have started their transmission simultaneously, a collision will be occurred and detected by the receiver. The PCTL property formulation is given in Equation 9, where ($ssi$=2) and ($str$=3) stand for SEND_RTS state of the sender i module and the COLLISION state of the receiver module respectively. PRISM checking of this property has returned true value.

$$P_{>=1}[true\,U\,(ss1 = 2)\&(ss2 = 2) => (str = 3)] \qquad (9)$$

In order to verify the effect of the chosen $T_{CU}$ duration in reducing the collision between hidden nodes, we have been studied the case where the senders have chosen a different values of backoff counter number rbc. In this case, we would verify that when the backoff counter *rbc* of the one of these senders reaches zero, the other must be detect activity in the channel and abort its idle listening in the current $T_{CU}$. The PCTL formulation of this property is given by the Equation 10, where ($ss2$=6) refers to the SLEEP state in *sender2* module, $b_{si}$ is a local variable that stores the value of backoff counter for the sender i model, and k is a constant in the range [1..12]. When this property is checked for each value of the constant k, PRISM model checker returns true value.

$$P_{>=1}[true\,U\,(bs1 = 0)\&(bs2 = k) \\ => (bs2 = k)\&(ss2 = 6)] \quad with \ \ k = 1..12 \qquad (10)$$

## 5. SENDER NEIGHBOR NUMBER VARIATION EFFECT

In order to assess the effect of varying the neighbor number of a given nodes on the successfully sending packet probability, we have written the PCTL quantitative property given in Equation 11, where ($ss1$=7) stands for the S_SUCC state in the PRISM *sender1* module, $es1$ is a local variable that stores the value of *e* for the module, and k represents a constants in the range [0..12]. We have evaluated this probability for different sender neighbor number. Each curve presented in figure 7 shows the variation of the successful sending probability related to a number of neighbor sender number. Each value in these curves evaluates the cumulative value of the probability returned by the Equation 11.

$$P_{=?}[F\,(ss1 = 7)\&(es1 = k)] \quad with \ \ k = 0..12 \qquad (11)$$

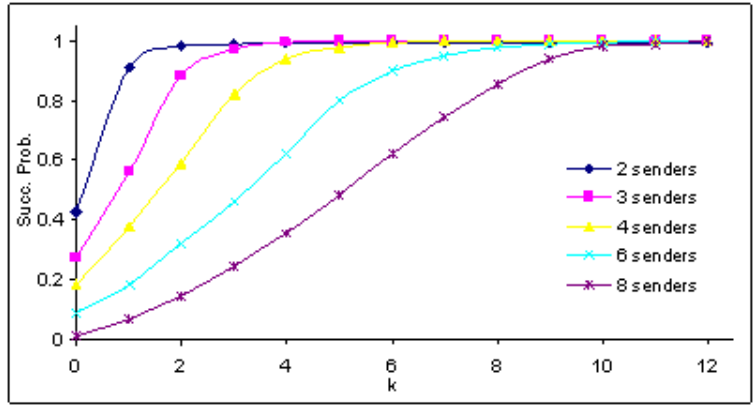

Figure 7. Success probability according to the unsuccessful transmission number in the past and relating to sender neighbor number

Results presented by figure 7, allow us to notice the following observations. (1) If the number of the sender is not sufficiently height, sender has more chance to send successfully its packet after a lowest number of $e$ that represents the number of unsuccessful transmissions in the past. (2) There is not much likelihood of the sender sending successfully its packet, if the number of senders is near to the contention window length $b_{max}$. We can conclude that the number of simultaneous sender toward a neighbor node in the network is related with the contention window length. This conclusion limits the simultaneous senders in the neighborhood, but it not restricts the neighbor number in the network. In fact, majority of WSN application is characterized by a light traffic load. These observations allow us to justify chosen bounds during the OPNET simulation phase. In fact, majority of simulated topologies are base on a grid structure where a non border node has eight neighbors.

## 6. CONTENTION UNIT LENGTH VARIATION EFFECT

This section studies the effect of varying the contention unit length $T_{CU}$ on the consumed energy introduced by the backoff procedure in idle listening state. This variation consists in increasing and decreasing the contention unit length against the value calculated using Equation 2 that we called initial value of $T_{CU}$.

### 6.1. Increasing the $T_{CU}$ value

In this section, we want to study the effect of increasing the contention unit duration $T_{CU}$ on the consumed energy dissipated in idle listening state. For that purpose, we chose to increase the $T_{CU}$ value by the one duration of $T_{FrmCtrl}$ that corresponds to the sending RTS control frame time. In that case, we use Rene sensor mote that consume 13.5 mW in idle listening state, we exploit Equation 4 for corresponding new $T_{CU}$ value and the property given in Equation 5, to evaluate dissipated energy in idle listening state of sender node. Figure 8 shows the dissipated energy for different values of the NMAX_MSG parameter using PRISM model checker based on a topology with two senders. These results are illustrated for the initial and increased $T_{CU}$ value. Results presented in figure 8, demonstrate that the increased value provides more dissipation of energy compared to the initial value of $T_{CU}$.

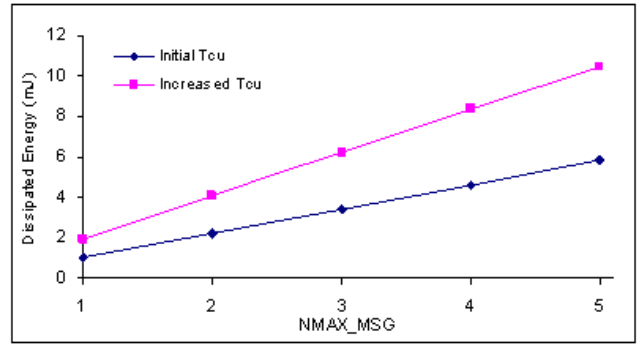

Figure 8. Dissipated energy for the initial and increased $T_{CU}$ values

## 6.2. Decreasing the $T_{CU}$ value

In order to study the effect of the $T_{CU}$ reduction, we have chosen to decrease the $T_{CU}$ value by the one duration of $T_{FrmCtrl}$ that corresponds to the sending RTS control frame time. After that, we have attempted to build the new model using PRISM model checker tool. In the building operation, PRISM displays several deadlock states in the new model. Since the PRISM tool offers a debugging interface based on a simulation engine guided by a user, we have decided to track an example of deadlock scenario. Figure 9 illustrates the trace of this simulation that is terminated with a deadlock state. In this example *sender1* and *sender2* have chosen two different values of *rbc* as follows 1 (column *sender1/bs1*) and 2 (column *sender2/bs2*) respectively. Consequently, the *sender1*'s *rbc* reaches zero before that of the *sender2*, and then sender1 switches from RX to TX (column sender1/ss1=1) to send its RTS (column *sender1/ss1*=2). The receiver follows the sender1 and it starts receiving the RTS (column *receiver/str*=1). At the end of sending, the sender1 switches from TX to RX (column sender1/ss1=3) for waiting CTS (column *sender1/ss1*=5) from the receiver that must switches from RX to TX (column *receiver/str*=4). We notice that the *sender2*'s *rbc* has reached zero and the sender2 starts switching from RX to TX for sending its RTS to the receiver that has terminated receiving *sender1*'s RTS and it switches to TX for sending CTS (column *receiver/str*=5) toward *sender1*. When the *sender2* starts sending its RTS, PRISM simulator engine displays the deadlock state.

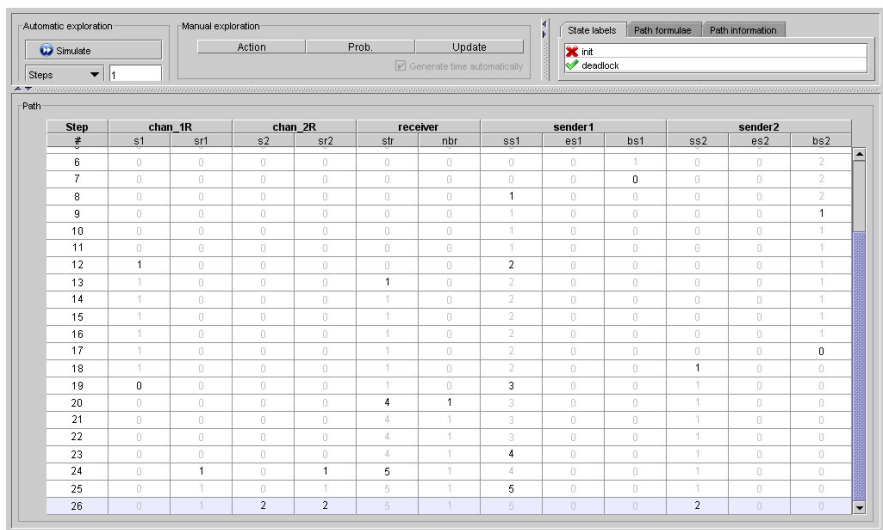

Figure 9. A deadlock scenario trace using PRISM's simulator engine

The deadlock state analyses allow us to notice that even if the senders have chosen different values of the random backoff counter *rbc*, it's likelihood to find new collision. In fact, the *sender2* has chosen a different *rbc* value and its transmission is failed because it has a collision on its RTS. In that case, the *sender2* increases its idle listening and the sending of the RTS is useless. These observations demonstrate that the decreasing of $T_{CU}$ value will increase the dissipated energy in idle listening and in useless sending operations.

Results obtained while increasing and decreasing the contention unit $T_{CU}$ length compared to the initial value, provide a rising of the dissipated energy. Therefore, these observations allow us to justify the initial used value of the $T_{CU}$.

## 7. CONCLUSION

In this paper, we have revealed that probabilistic model checking can be used to formally verify interesting properties of contention backoff procedure that would be difficult to discover using alternative simulation analysis techniques. The backoff procedure is used in contention phase of the wireless sensor network ECo-MAC [7] protocol implemented in OPNET simulator. For that, we started by a description of the main works that have contributed to modeling and analyzing wireless communication protocols using probabilistic model checking technique that stands more by modeling probabilistic behaviors. Next, we have presented our proposed backoff procedure model using probabilistic model checker PRISM framework and also we have proved that the model is conform with that implemented in OPNET simulator using comparison between results obtained in the both OPNET and PRISM tools. We have used the rewards concept to find results in PRISM. Then, we have used the PCTL logic language to formulates different properties allow us to ensure that the proposed model performs their basic functionality. Also, we have used the quantitative property to study the effect of the contention window length on the number of the simultaneous senders in a same neighborhood. We have found that the bound is near of the window length. Finally we have justified the good choice of the contention unit $T_{CU}$ value that is made in OPNET simulation phase. This choice provides the optimal dissipated energy for the sender node.

We can extend this work by the following ideas. In the first, we will introduce more reduction in the proposed model. The second idea consists in modeling the all specification of the WSN ECo-MAC protocol. Another direction would be to combines PRISM and UPPAAL modeling and analyses of larger network configurations.

**Authors**

Zayani Hafedh received his Engineer Diploma in Computer Science from Tunisia national engineering school ENSI in 2001; his Master Diploma in Telecommunication System (SysCom) from Tunisia national engineering school ENIT in 2003 and his Ph.D. degree in computer science and telecommunication from ENI-Tunis&CNAM-Paris in 2009. He is working at SysCom Research Laboratory of ENIT. His research is focused on wireless sensors networks performance optimization, focused in the communication protocols.

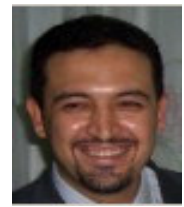

Kamel Barkaoui received a Ph.D. degree in Computer Science in 1988 from the University Paris 6. He is currently a Professor at the Conservatoire National des Arts et Métiers (CNAM - Paris). His research interests include verification techniques and performance evaluation methods of concurrent and distributed systems and their applications to computer and communication systems. Dr Barkaoui has served on PCs and as PC chair for numerous international workshops and conferences. He was a guest editor of

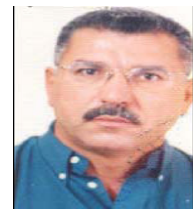

Journal of Systems and Software and has received the 1995 IEEE Int. Conf. on System Man and Cybernetics Outstanding Paper Award.

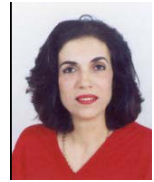

Rahma Ben Ayed received her Doctorat in Computer Science from the Faculty of Science of Tunis in 1990. She is currently a Professor at the Department of Information and Communication Technologies at the Ecole Nationale d'Ingénieurs de Tunis (ENIT) and a member of SysCom Research Laboratory. Her research interests include V&V of software systems, dependability and wireless sensor networks. She teaches several courses on programming languages, software engineering and protocol engineering.